
\def\dslash{\partial}
\font\titlefont=cmbx10 scaled\magstep1
\magnification=\magstep1
\null
\rightline{SISSA 45/94/EP}
\vskip 1.5cm
\centerline{\titlefont RENORMALIZATION GROUP EFFECTS}
\centerline{\titlefont IN THE CONFORMAL SECTOR}
\centerline{\titlefont  OF 4D QUANTUM GRAVITY WITH MATTER}
\smallskip
\vskip 1.5cm
\centerline{\bf S. Odintsov \footnote{$^*$}{\tt
odintsov@ebubecm1.bitnet}}
\smallskip
\centerline{Dept. ECM, Barcelona University}
\centerline{Diagonal 647, 08028 Barcelona, Spain}
\bigskip\smallskip
\centerline{\bf R. Percacci \footnote{$^{**}$}{\tt
percacci@tsmi19.sissa.it}}
\smallskip
\centerline{International School for Advanced Studies, Trieste,
Italy}
\centerline{via Beirut 4, 34014 Trieste, Italy}
\centerline{and}
\centerline{Istituto Nazionale di Fisica Nucleare,
Sezione di Trieste}
\vskip 1.8cm
\centerline{\bf Abstract}
\smallskip
\midinsert
\narrower\narrower\noindent
We discuss the ``gravitationally dressed'' beta functions in the
Gross--Neveu model interacting with 2d Liouville theory and in $SU(N)$
gauge theory interacting with the conformal sector of 4d quantum
gravity. Among the effects that we suggest may feel the gravitational
dressing are the minimum of the effective potential and the running of
the gauge coupling.
\endinsert
\vskip 1cm
\vfil\eject

The influence of quantum gravitational effects on the renormalization
group (RG) is expected to be very complicated. Already in the case of
a quantum field theory in curved spacetime, the RG is getting quite
nontrivial [1]. For example, many new couplings appear which are
absent in flat space.

Recently it has been discovered [2] that 2d quantum gravity gives
a ``gravitational dressing'' of the matter beta functions. Then, the
RG equations of matter are modified even in the situation when the
divergences are formally the same as in the absence of quantum gravity
[2,3,4].

In the present note we try to find a 4d analog of the ``gravitational
dressing'' of the matter beta function using the effective theory of
the conformal factor [5] as an example.
We shall begin by giving an alternative, RG based, derivation of this
theory. We then consider the RG dressing of the beta function in the
case of the 2d Gross--Neveu model coupled to gravity.
Finally we apply the same method to 4d $SU(N)$ gauge theory coupled
to the conformal sector of Quantum Gravity (QG).

We begin by considering a massive self-interacting scalar theory in curved
spacetime. Our first purpose is to show that the conformal sector of
quantum gravity can be fixed by the requirement of multiplicative
renormalizability of such a theory, with the metric treated as an
external field. Let us therefore start from the action
$$
S(\varphi)=-\int d^4x\ \sqrt{g}
\left[{1\over2}g^{\mu\nu}\partial_\mu\varphi\partial_\nu\varphi
+{1\over2}(m^2+\xi R)\varphi^2+{f\over 4!}\varphi^4\right]\ ,
\eqno(1)
$$
In what follows we will restrict our attention to metrics of the form
$$
g_{\mu\nu}=e^{2\sigma}\eta_{\mu\nu}\ ,\eqno(2)
$$
where $\eta_{\mu\nu}$ is the flat metric and $\sigma$ is the
conformal factor, which will be the only dynamical component of the
gravitational field. Defining
$$
\phi=e^\sigma\varphi\ ,\eqno(3)
$$
the action (1) can be written as
$$
S(\sigma,\rho)=-\int d^4x\
\Bigl[{1\over2}\eta^{\mu\nu}\partial_\mu\phi\partial_\nu\phi
+{1\over2}m^2 e^{2\sigma}\phi^2+{f\over 4!}\phi^4
+{1\over2}\tau\phi^2(\partial^2\sigma
+\eta^{\mu\nu}\partial_\mu\sigma\partial_\nu\sigma)\Bigr]\ ,
\eqno(4)
$$
where $\tau=1-6\xi$. Note some nonstandard terms describing the
interaction of the scalar field $\phi$ with $\sigma$. All these terms
disappear for $m^2=0$, $\tau=0$, {\it i.e.} when (1) is invariant
under Weyl transformations of $g_{\mu\nu}$.

We are going to study the
renormalization of the theory (4) regarded as a quantum theory in flat
space coupled to a scalar background field $\sigma$.
According to the standard renormalization formalism in a background
field (see for example [1]) in order to guarantee the multiplicative
renormalizability of the theory one has to add to the action (4) the
action of the external fields.
Considerations of dimensionality, covariance and explicit loop
calculations fix the external field action to be of the form
$$
S_{\rm ext}=\int d^4x\,\left[-{Q^2\over(4\pi)^2}(\partial^2\sigma)^2
-\zeta\left[2(\partial_\mu\sigma)^2\partial^2\sigma+(\partial\sigma)^4\right]
+\gamma e^{2\sigma}(\partial_\mu\sigma)^2
-\lambda e^{4\sigma}\right]\ .\eqno(5)
$$
This is exactly the same action that one gets as a result of
integrating over the conformal anomaly [7,5].  (For a recent very
interesting discussion of the general structure of the conformal
anomaly in $d$ dimensions, see [8]).
In [5] the coupling constants of (5) were fixed in terms of the
coefficients of the conformal anomaly, and $S_{\rm ext}$ was used to
describe the IR region of QG.
We have a completely different interpretation of $S_{\rm eff}$ --
its form is required by the condition of multiplicative renormalizability.
The effective potential for the conformal factor in the theory (4)
was discussed in [6].

Let us discuss now the renormalization of the theory under
discussion. By the standard calculation of one loop divergences,
one can find the beta functions, and from there the effective running
coupling constants
\smallskip
\settabs\+\indent\qquad\qquad&$\zeta(t)=\zeta+{\tau^2\over2f}B(t)$\qquad\qquad
&\cr
\+&$f(t)={f\over A(t)}$&$m^2(t)=m^2 A(t)^{-1/3}$\cr
\+&$\tau(t)=\tau  A(t)^{-1/3}$&\cr
\+&$\zeta(t)=\zeta+{\tau^2\over2f}B(t)$&
$Q^2(t)=Q^2+{(4\pi)^2\over2f}\tau^2B(t)$\cr
\+&$\gamma(t)=\gamma+{m^2\tau\over f}B(t)$
&$\lambda(t)=\lambda+{m^4\over2f}B(t)$\cr
\smallskip
\noindent
where $A(t)=1-{3f\over(4\pi)^2}t$ and $B(t)=A(t)^{1/3}-1$.
One sees that both in the IR limit $t\rightarrow-\infty$,
where the theory is asymptotically free, and in the UV
limit, where the problem of zero charge appears, the effective
couplings tend to their conformally invariant values.
At the same time the couplings of $S_{\rm ext}$ grow
infinitely with $|t|$.

We are interested in the physical scaling of 4d IR quantum gravity
with matter, {\it i.e.} the theory with action (5)
where $\sigma$ is also a quantum field now.

Before doing this, we discuss first the analogous situation in $d=2$,
namely the case of Liouville theory.
The effective Lagrangian for the Liouville part of the induced
quantum gravity may be written
$$
L=-{Q^2\over 4\pi}(\nabla\sigma)^2-2\lambda e^{2\sigma}\ ,\eqno(6)
$$
where the conformal gauge $g_{\mu\nu}=e^{2\sigma}\eta_{\mu\nu}$
has been used, $Q^2={1\over12}(25-c)$ and $c$ is  the central charge.
Quantizing $\sigma$ one finds that there
is an anomalous scaling behaviour. It can be defined by saying
that the conformal factor $e^\sigma$ acquires a scaling dimension
$\alpha$ ({\it i.e.} putting $\hat\sigma=\alpha\sigma$),
calculating the exact beta function for $\lambda$ and demanding
the vanishing of this beta function for $\lambda\neq 0$.
This gives [9]
$$
\alpha={1-\sqrt{1-{2\over Q^2}}\over  {1\over Q^2}}\ ,\eqno(7)
$$
where the negative branch of the square root is to be chosen.
The classical scaling $\alpha=1$ is obtained in the limit
$Q^2\rightarrow\infty$ (no quantum gravity).
Substituting $Q^2$ in (7) one has
$$
\alpha=Q\left(Q-\sqrt{Q^2-2}\right)=
\sqrt{25-c\over12}\left(\sqrt{25-c\over12}-\sqrt{1-c\over12}\right)
\eqno(8)
$$
showing that the critical exponent is real only for $Q^2>2$.
In terms of the critical exponent, the renormalized dimensionless
cosmological constant satisfies the exact renormalization group
equation [3,10]
$$
{\partial\lambda\over\partial\log\mu}=-2\alpha\lambda\eqno(9)
$$
with the solution
$$
\lambda(\mu)\sim\mu^{-2\alpha}\ .\eqno(10)
$$
If $c\rightarrow-\infty$ (which corresponds to the absence of quantum
gravity), $\lambda(\mu)\rightarrow\mu^{-2}$.

Now let us add to the 2d induced quantum gravity a matter action
corresponding for example to the Gross--Neveu model [11]
(The interaction of this model with dilaton gravity has been
considered recently in [12]). The Lagrangian is given by
$$
L=\bar\psi(i\dslash)\psi-{1\over2}\sigma^2-g\sigma\bar\psi\psi\
,\eqno(11)
$$
where $\sigma$ is a scalar and $\psi$ is an $N$ component massless
fermion.
In the leading order of the $1/N$ approximation, the beta function
of this theory does not change in the presence of
gravity and is given by
$$
\beta_g=-{\lambda g\over2\pi},\qquad
\lambda=Ng^2\ .\eqno(12)
$$
The running coupling constant has the asymptotically free form
$g^2(t)={g^2\over\left(1+{\lambda t\over\pi}\right)}$.
The gravitational dressing of the beta function (12), can be derived
by the following argument [3]. In the presence of quantum gravity the
physically meaningful beta functions are not the derivatives of the
coupling constants with respect to the mass scale $\mu$, but rather
the derivatives with respect to some other scale coming from the
gravitational sector [10]. In [3] the cosmological term was
chosen, but the result is universal. Then
$$
\beta^G_g={\partial g\over\partial\log\lambda^{-1/2}}=
{\partial g\over\partial\log\mu}
{\partial\log\mu\over\partial\log\lambda^{-1/2}}=
{1\over\alpha}\beta_g\ .\eqno(13)
$$
This is the ``gravitational dressing'' of the beta function [2,3,4].
It is interesting to understand how these considerations explicitly
modify the running coupling constant and effective potential.
{}From (13) we get
$$
{dg(t)\over dt}=-{\lambda(t)g(t)\over2\pi\alpha} \eqno(14)
$$
and the ``gravitationally dressed'' running coupling is
$g^2(t)={g^2\over\left(1+{\lambda t\over\pi\alpha}\right)}$.
Hence the value of the pole in the IR region changes from
$t=-{\pi\over\lambda}$ (no QG) to $t=-{\pi\alpha\over\lambda}$
(with QG). Now let us consider the one loop effective potential
in this theory. It may be found as a solution of the RG equation [13]
$$
\left[\mu{\partial\over\partial\mu}
+\tilde\beta^G{\partial\over\partial g}
-\tilde\gamma^G\sigma{\partial\over\partial\sigma}\right]V=0\ .
\eqno(15)
$$
Here
$$
\tilde\beta^G={1\over 1-\gamma^G}\beta^G\ ,\qquad
\tilde\gamma^G={1\over 1-\gamma^G}\gamma^G\ ,\eqno(16)
$$
as usual. Note that in these equations we have to use the
gravitationally dressed RG functions.
We can solve the RG equation (15) and find
$$
V={1\over2}\sigma^2+{\lambda\over4\pi\alpha}
\sigma^2\left[\log{\sigma^2\over\mu^2}-3\right]\ .\eqno(17)
$$
The effect of the gravitational dressing  is the factor $1/\alpha$
in the second term. In principle, it could lead to some observable
physical effect. In fact, the minimum of the potential
and the fermionic mass are given by
$$
\sigma_m=\mu\exp\left(1-{\pi\alpha\over\lambda}\right)\ ,
\qquad M_f=g\sigma_m\ .\eqno(18)
$$
Both these values feel the effect of gravity.
Of course these considerations are somehow formal
(just as in [2,3,4]) because in this model  $c>1$ (for $N>1$). However,
in principle, one could add to the system some exotic free conformal field
theory to lower the value of $c$ without changing the dynamics. In any case,
this model gives some indication of what one can expect to find in more
realistic situations, like the d=4 case that we now return to.

We work within the framework of the effective dynamics for the conformal
sector which was discussed first in [5], as a 4d analog of Liouville theory
and was rederived from a different viewpoint in the beginning of this paper.
We consider the effective action (5) for the conformal factor in the IR
stable fixed point $\zeta=0$, which presumably corresponds to the IR sector
of QG. The Lagrangian is therefore
$$
L=-{Q^2\over(4\pi)^2}(\partial^2\sigma)^2
+\gamma e^{2\sigma}(\partial_\mu\sigma)^2
-\lambda e^{4\sigma}\ .\eqno(19)
$$
where $Q^2=-32\pi^2 b'$, and  $b'$ is the coefficient of the Gauss-Bonnet
term in the conformal anomaly.
Proceeding from here as in the two dimensional case one can put
$\sigma=\alpha\hat\sigma$, find the exact beta functions for $\gamma$ and
$\lambda$ and from the condition of vanishing beta functions derive the
anomalous scaling dimension of $e^\sigma$ [5]:
$$
\alpha={1-\sqrt{1-{4\over Q^2}}\over  {2\over Q^2}}\ ,\eqno(20)
$$
This critical exponent is real for $Q^2>4$, which is the standard
situation in $d=4$.

In terms of this critical exponent we have again an exact RG equation
for the renormalized cosmological constant
$$
{\partial\lambda\over\partial\log\mu}=-4\alpha\lambda\eqno(21)
$$
giving $\lambda\sim\mu^{-4\alpha}$. In the limit of no QG,
$Q^2\rightarrow \infty$, $\alpha\rightarrow 1$.

By analogy with the 2d case we define the physical mass scale
as $\lambda^{-1/4}$ (the same results would be obtained if we used
instead $\gamma^{-1/2}$).
Let us now couple the theory with Lagrangian (5) to
some massless matter theory consisting, for simplicity,
of spinors and $SU(N)$ gauge fields.
We avoid scalar and Yukawa couplings in this context, since their
beta functions are known to receive contributions from the
spin 2 sector of quantum gravity (for an example
in the context of $R^2$--gravity see [15,1]).
On the contrary, the beta function of the gauge coupling does not change
in the presence of gravity [14,15].
Hence, we expect that transverse spin 2 degrees
of freedom may be safely neglected in this case.

Writing the Lagrangian of spinor and gauge fields in an external metric of
the form (2) and making the appropriate rescalings of the fields as in (3),
one can see that this theory does not couple to the scalar $\sigma$.
Hence the one-loop beta function for the gauge coupling is the same as
in flat space:
$$
\beta_{g^2}={\partial g^2\over\partial\log\mu}=-a^2g^4\ ,\eqno(22)
$$
where the constant $a^2$ depends on the details of the theory.

As in 2d the effect of QG will be to replace the beta function (22)
by the gravitationally dressed beta function
$$
\beta^G_{g^2}={\partial g^2\over\partial\lambda^{-1/4}}=
{\partial g^2\over\partial\log\mu}
{\partial\log\mu\over\partial\log\lambda^{-1/4}}=
{1\over\alpha}\beta_{g^2}\ .\eqno(23)
$$
The gravitationally dressed running gauge coupling is
$$
g(t)^2={g^2\over\left(1+{a^2g^2t\over\alpha}\right)}\ .\eqno(24)
$$
For an $SU(N)$ model, taking into account also the contributions to
$Q^2$ coming from Einstein gravity or Weyl gravity, and from the
$\sigma$ sector, we have
$$
Q^2={1\over180}\left(11N_F+62N_V+n\right)\ ,\qquad
a^2={1\over(4\pi)^2}\left[{22\over3}N-{4\over3}
\sum_{\rm F}T(R)\right]\ ,\eqno(25)
$$
where the sum in the last equation is over all fermion representations
and $n=1383$ for Einstein gravity, $n=1583$ for Weyl gravity [16].
For example in the case of the standard model one has
$\alpha=1.13$ and $1.11$ respectively. We are neglecting here the
scalar contribution to $Q^2$, which is very small with respect to the
rest. As we see, the gravitational dressing of the gauge coupling beta
function has quite modest effects, of the order of few percents.
Taking this effect into account in the running of the gauge coupling
constants $\alpha_1$, $\alpha_2$, $\alpha_3$ corresponding to the
groups $U(1)$, $SU(2)$ and $SU(3)$, and imposing that they meet at
some unification scale, we find an increase in this scale of a factor
of the order 2, compared to the case in which gravity is neglected.
This is in principle an observable effect.

We would like to conclude with a remark  on the effective potential in
the $SU(N)$ gauge theory. Consider the composite field
$K=A_\mu A^\mu$. The effective potential for $K$ has been calculated in
[17]:
$$
V=-{3n\over(4\pi)^2\alpha}K^2\log{K\over\mu^2}\ ,\eqno(26)
$$
in the Landau gauge ($n$ is the dimension of the group).
Due to the gravitational dressing, this
potential is modified by the appearance of a factor
$1/\alpha$. As in the case of the Gross--Neveu model,
this may lead to a modification of the dynamically generated mass,
which was discussed in [17] in the case $\alpha=1$.

\bigskip
\centerline{\bf References}
\item{1.} I.L. Buchbinder, S.D. Odintsov and I.L. Shapiro,
``Effective action in Quantum Gravity'', IOP publishing,
Bristol and Philadelphia (1992).
\item{2.} I.R. Klebanov, I.I. Kogan and A.M. Polyakov,
Phys. Rev. Lett. {\bf 71} (1993) 3243.
\item{3.} Y. Tanii, S. Kojima and N. Sakai, Phys. Lett. {\bf B 322}
(1994) 59.
\item{4.} J. Ambj\o rn and K. Ghoroku, NBI HE 93-63 (hep-th 9312002).
\item{5.} I. Antoniadis and E. Mottola, Phys. Rev. {\bf D 45} (1992)
2013.
\item{6.} R. Floreanini and R. Percacci, SISSA 71/93/EP (hep-th
9305172)
\item{7.} R. Riegert, Phys. Lett. {\bf B 134} (1984) 56;\hfil\break
E.S. Fradkin and A.A. Tseytlin, Phys. Lett. {\bf B 134} (1984) 187;
\hfil\break
I.L. Buchbinder, S.D. Odintsov and I.L. Shapiro, Phys. Lett. {\bf B
162} (1985) 93;\hfil\break
E.T. Tomboulis, Nucl. Phys. {\bf B 329}, (1990) 410;\hfil\break
S.D. Odintsov and I.L. Shapiro, Class. and Quantum Grav. {\bf 8}
(1991) L57;\hfil\break
I. Antoniadis and S.D. Odintsov, Mod. Phys. Lett. {\bf A 8} (1993)
979.
\item{8.} S. Deser and A. Schwimmer, Phys. Lett. {\bf B 309} (1993)
279.
\item{9.} A.M. Polyakov, Mod. Phys. Lett. {\bf A 2} (1987)
893;\hfil\break
F. David, Mod. Phys. Lett. {\bf A 3} (1988) 1651;\hfil\break
J. Distler and H. Kawai, Nucl. Phys. {\bf B 321} (1989) 509.
\item{10.} H. Kawai and M. Ninomiya, Nucl. Phys. {\bf B 336} (1990)
115;\hfil\break
H. Kawai, Y. Kitazawa and M. Ninomiya, Nucl. Phys. {\bf B 404} (1993)
684.
\item{11.} D.J. Gross and A. Neveu, Phys. Rev. {\bf D 10} (1974) 3235.
\item{12.} T. Muta, S.D. Odintsov and H. Sato, Mod. Phys. Lett.
{\bf A 7} (1992) 3765.
\item{13.} S. Coleman and E. Weinberg, Phys. Rev. {\bf D 7} (1973)
1888.
\item{14.} S. Deser, H.S. Tsao and P. van Nieuwenhuizen,
Phys. rev. {\bf D 10} (1974) 3337.
\item{15.} E.S. Fradkin and A.A. Tseytlin, Nucl. Phys. {\bf B 201}
(1982) 469.
\item{16.} I. Antoniadis, P. Mazur and E. Mottola, Nucl. Phys. {\bf B 388},
627 (1992);\hfil\break
S.D. Odintsov, Z. Phys. C {\bf 54}, 531 (1992).
\item {17.} R.P. Grigoryan and I.V. Tyutin, Sov. J. Nucl. Phys.
{\bf 26} (1977) 593.

\bye